*Biomimetic random lasers with tunable spatial and temporal coherence*

*Neda Ghofraniha\*, Luca La Volpe, Daniel Van Opdenbosch, Cordt Zollfrank and Claudio Conti*

Dr. N. Ghofraniha, L. La Volpe, Prof. C. Conti
Institute for Complex-Systems CNR, UOS Sapienza,
University La Sapienza, P.le A. Moro 5, 00185, Roma, Italy.

Department of Physics, University La Sapienza, P.le A. Moro 5, 00185, Roma, Italy

E-mail: neda.ghofraniha@roma1.infn.it

Prof. D. Opdenbosch, Prof. C. Zollfrank

Chair of Biogenic Polymers, Technische Universität München (TUM) at the Straubing Center of Science for Renewable Resource , Schulgasse 16, 94315 Straubing, Germany

Living organisms have evolved well-adapted structures and materials over geological eras. Through evolutionary selection, nature has devised effective solutions to all sorts of complicated real-world problems and, following Leonardo Da Vinci, humans have looked at nature to reach answers.  The young field of biomimetics has given rise to new technologies inspired by  nature's strategy for materials and devices optimized from the macroscale to the nanoscale.

The complex interplay between surface morphology and physical and chemical characteristics is responsible for the properties of biological materials. Superhydrophobicity, self-cleaning, energy conversion and conservation, reversible large adhesion, biological self-assembly, antireflection, thermal insulation, self-healing and sensory-aid mechanisms are some of the examples found in nature that are of commercial interest. Hierarchical structures with dimensions ranging from the macroscale to the nanoscale   continuously inspire novel routes for advanced functional materials with applications in  material technology[1-9] and photonics.[10-16]

Recently, cellulose has been proposed as a new material for photonic crystals and random structures. Materials derived from cellulose provide coloration and ultra-white response for applications as security labeling or artificial pigments.[17,18] The disordered arrangement of cellulose fibers in paper has been used as optical cavity for versatile organic Random Lasers (RLs), where the emission of a gain medium is amplified by multiple scattering inside the paper mesh. The RL emission has been identified as a wide spectral peak over the gain fluorescence due to the incoherent feedback mechanism.[19,20] Random lasers have been realized and analyzed by several research groups in the last two decades.[21-23] Random lasing in biomimetic structures withß emission intensity controlled by the structure of the material have been reported[15] and very recently a bio-inspired RL with quasi-periodic structure of butterfly wings demonstrate quasi single mode emission in presence of Fabry-Perot resonator[24].

For a variety of possible applications, we lack technology for biomimetic lasers with tunable wavelength, longitudinal and transverse coherence. The spatial extension of the modes has been studied in standard RLs by different techniques.[25-29] Such methods were limited to few microns emission area in systems under selective pumping and with localized modes that could be switched and interact. Emission from aggregates at long distances has been also reported but without measuring the spatial extension of each single mode.[30]

The novelty of our work is to tune the interaction of few resonant modes with wide spatial length by mapping their emission intensity separately and changing suitably the pumping parameters.

Here we realize novel random laser devices made by biotemplated paper, where titania nano-particles with 2.3 refractive index replace cellulose, that has index of refraction 1.5, and strongly enhance light scattering. The enhancement provides efficient RL with respect to similar devices made by standard paper. The scattering is so strong that we find stable distinct modes with sharp sub nanometric spectral line width and with unexpected broad spatial emission. Indeed we estimated the mean free path in pure paper equal to 37 μm that is more than twice of that of the biotemplated one, as detailed below. We adopt an advanced spectral mapping technique able to pattern for the first time the RL emission on large lengths (up to 900 μm$^2$). Indeed, we observe modes with localization lengths in a range of 10-25 μm and on the detectable large areas there is clear evidence for their long-range nonlinear interaction and competition.

This nonlinear regime is the key for the control of the emission frequency, line width and spatial pattern of the desired operational modes by acting on the pumping parameters as the energy and the shape of the pump.

The biomimetic samples are sections of titanium dioxide ($TiO_2$) calcinated paper of thickness (100 ± 20)μm and mean free path (17 ± 4)μm, as estimated by transmission measurements. Details on the preparation of the biotemplated paper are reported in Experimental Section.

After calcination of the infiltrated round filter papers, white sheets that are shrunk by 30 % in diameter are obtained. They can be handled with ease and cut with scissors for analysis, already hinting at their porous, but interwoven microstructure. Scanning electron micrographs (SEM), with an example shown in **figure 1**a, confirm their randomly oriented fibrous microstructure, which is retained from the filter paper template. The X-ray diffractograms of native and infiltrated filter papers, figure 1b (I) and (II), show reflexes from cellulose. Additionally, reflexes from quartz are detected and interpreted as a portion of the naturally occurring ash. The diffractogram of the infiltrated paper shows no reflexes additional to the native paper. However, after calcination, figure 1b (III), broad reflection peaks from anatase are present. This indicates a structure formation of crystalline titania from a condensed, but still amorphous, TEOT precursor during calcination. This is conclusive, since the condensation of alkoxide precursors progresses via hydrolysis under loss of alcohol and condensation under loss of water at ambient conditions. In a related study, tetraethyl orthosilicate (TEOS) in a similar cellulosic template was found to be hydrated and weakly coordinated.[31] After calcination, a dry material with mainly three and fourfold coordination is obtained after firing at 500 °C. In this study using titania as the target phase, expectedly, a crystallized material is obtained after the thermal template removal. The crystallite size is estimated as (8 ± 4) nm. Such small crystal sizes are typically obtained from TEOT-based precursor solutions.[32,33] In our work, this means that the crystal domains are small enough to provide a detailed replica of the paper template, confirming our observations by SEM.

To realize the RLs the so obtained biotemplated paper sections are placed in a 3 mM solution of rhodamine B in diethylene glycol acting as gain medium and the dyed dispersion is then put in a transparent 250 μm thick sealed chamber and illuminated by the laser stripe as shown in **figure 2**b.

The experimental setup for the detection of the RL modes is reported in figure 2a: a Q-switched Nd:Yag pulsed laser pump the material and the emission, with a magnification of 100X, is contemporary visualized on a charge coupled device (CCD) camera and spectrally mapped in space by moving a fiber along the X-Y directions with 0.5 μm resolution. The diameter waist $W_P$ of the laser stripe along the Y axis varies in the range of 16-36 μm, details are reported in Experimental Section. We scan the spectral emission with sub micrometric resolution along two directions and analyze the so obtained intensity maps to show the variation of mode composition point by point by changing pump energy and width.

We report in figure 2c a fluorescence image of one dyed TiO2 calcinated paper section pumped at $E_P$=6.4 μJ energy as displayed on the CCD and in figure 2d the emission map from whole the sample: the intensity of each pixel is given by the integral of the emission spectrum detected by the fiber at the corresponding position.
The obtained 30 μm X 30 μm intensity spectral map reproduces quite accurately the CCD image, giving quantitative information on the spatial extension of the emission.
The fiber scanning technique allows the detection of the RL emission and the composition of modes point by point and our findings are reproducible in a variety of different samples.
In the following, we report on results on three representative samples named α, β and γ. Each device differs in the spectral features, but the mode competition dynamics discussed below is common of all the considered cases.
In **figure 3** we report results for sample α. Figures 3a and figure 3b illustrate 13.5 μm X 13.5 μm intensity maps of two resonances at respectively $\lambda_1^\alpha$=594.54 nm and $\lambda_2^\alpha$=600.26 nm, as evidenced by the two representative spectra in figure 3c detected at the indicated points.
In these maps each pixel corresponds to the area of one single peak as depicted in green and red colors in figure 3c. Each map provides the visualization of the spatial extension of one single mode. The fact that the mode at $\lambda_2^\alpha$ =600.26 nm in figure 3b is confined at the edge of the RL can be ascribed to the growth of gain in the direction of loss boundary as demonstrated in.[34] From figure 3 it is also noticeable that the composition of modes change in space, evidencing the spatial coexistence of distinct modes only in some points of the system. The two analyzed sharp peaks at $\lambda_1^\alpha$=594.54 nm and $\lambda_2^\alpha$ =600.26 nm have line width $\Delta\lambda$ =0.2±0.1 nm.

The lasing behavior is demonstrated by the drastic increase in the intensity of the peak by increasing pumping energy $E_P$ as reported in figure 3d, where the two modes present two distant energy thresholds $E_{th}$ at respectively $E_P$ =6.55 µJ and $E_P$ =5.69 µJ, as estimated by the intersection of the linear curves fitting the data.

The dependence of mode composition and interaction on the input energy is shown in **figure 4** for sample β. This sample presents two RL resonances at $\lambda_1^\beta$=605 nm and $\lambda_2^\beta$ =606.05 nm, whose 15 µm X 15 µm intensity maps are reported in panels a-d at two different pumping energy $E_P$. At some spatial location one mode is dominant, while in others the two modes coexist. Moreover, their configuration changes both in intensity and in the spatial profile for different pumping energy. Figure 4e shows two spectra for the resonances at $\lambda_1^\beta$ and $\lambda_2^\beta$ at two different input energies. It is evident that they compete for the available energy that is transferred from mode $\lambda_2^\beta$ =606.05 nm to mode $\lambda_1^\beta$ =605 nm for increasing pumping.
These spectra correspond to the pixels where the two modes have the highest intensity and are normalized for a better visualization.

We quantify the effect of mode competition by estimating, for samples α, β and γ, the energy $E_k$ and the localization length[35] $l_k$ of single modes as

$$E_k^{\alpha,\beta,\gamma} = \frac{\sum_{i,j=1}^{N} I_{ij}^k}{\sum_{i,j=1}^{N} I_{ij}^{tot}} \qquad (1)$$

$$l_k^{\alpha,\beta,\gamma} = \sqrt{\frac{\left(\sum_{i,j=1}^{N} I_{ij}^k\right)^2 \Delta X_i \Delta Y_j}{\sum_{i,j=1}^{N} (I_{ij}^k)^2}} \qquad (2)$$

where k is the mode index, N is the total number of pixels, $I_{ij}^k$ is the intensity of the k-indexed resonance, calculated as the area below the peak in the spectrum at position ij along X-Y, and $I_{ij}^{tot}$ is the area of entire spectrum at position ij. $\Delta X_i$ and $\Delta Y_j$ are pixel lengths along the axis ($\Delta X_i = \Delta Y_i$=0.5 µm).

The variation of mode energy $E_k^\beta$ and spatial extension $l_k^\beta$ of sample $\beta$ with pump laser energy $E_P$ are illustrated in figure 4f and figure 4g for the two modes. Mode competition is clearly evident: by increasing energy, $\lambda_2^\beta$ loses energy and broadens in space in favor of $\lambda_1^\beta$. The spatial coherence of the two modes is so controlled by the pumping energy. In these measurements the beam diameter waist is $W_P = 15.1 \pm 0.4$ μm.

Sample $\gamma$ presents one single mode and in **figure 5** we report the dependence of its spatial extension and spectral configuration on the waist $W_P$ of the pumping beam. The normalized intensity maps of one mode reported in figure 5a-c visualize the enlargement of mode spatial extension for growing $W_P$, moreover the spectral configuration drastically change as shown in figure 5d, where the spectral line centered at $\lambda_1^\gamma = 603.7$ nm broadens, as the result of more overlapping broad resonances that emerge under the gain spectral profile by augmenting the pumped area.

We estimate the localization length $l_1^\gamma$ by using equation (2) where $I_{ij}^k$ is calculated as the area below the spectral peak in the range marked with dashed lines in figure 5d, the same used for the intensity maps in panels a-c, and the spectral line width $\Delta\lambda^\gamma$ as the full width at half maximum of the peak. Figure 5e shows the growing of $l_1^\gamma$ and $\Delta\lambda^\gamma$ for increasing $W_P$. For larger pumped area more modes are activated with broader coalescing emission lines and less localized is space. In this way by changing the pumping shape miniaturized lasers with variable temporal coherence ($\Delta\lambda^\gamma$) and spatial coherence ($l^k$) are obtained.

In conclusion, we fabricated disordered active materials mimicking the complex structure of paper. The random lasers realized by this material exhibit sharp distinct resonances with broad localization lengths. We used a new technique to directly visualize and quantify with sub micro metric resolution and for the first time on large area ($\approx 10^3$ μm$^2$) the spectral and spatial patterns of the modes. We showed that on these large length scales RL modes interact and compete for the available energy: By increasing pumping some modes gain intensity and reduce their localization length at the expense of others. This effect is strongly dependent on the energy and on the width of the input laser and allows high resolution control of the emission characteristics by finely tuning the pumping parameters. Our findings are of wide interest both in the fundamental research of the physics of random lasers and in the investigation and production of novel photonics functional materials and advanced miniaturized devices with applications in biomedicine and bio-imaging. The proposed nature-inspired active composite can be used as synthetic marker with intense spectral emission tunable by suitably designing its micro structure, and as a miniaturized source of light[36] with variable coherence.

**Experimental Section**

*Sample preparation:* Round cellulose filter papers (Type 15A, Rotilabo, VWR, Darmstadt, Germany) are vacuum- infiltrated at 100 mbar with a 10 mass% ethanolic solution of tetraethyl orthotitanate, TEOT (ethanol 99.8 %, denatured with methyletherketone, VWR).

The amount of TEOT is 1.41 g per g of filter paper. Ethanol is removed by drying at 50 °C. Residual condensation water is eliminated by curing at 105 °C. All organic substances are removed by calcination at 500 °C .To avoid warping during calcination, the papers are evenly weighed with 1 mm alumina plates.

The microstructures of the templates, infiltrated and calcined materials were determined by scanning electron microscopy (TUM machine). The phases of the calcined materials were determined via X-ray diffration (Miniflex 600, Rigaku, Tokyo, Japan) by pattern matching (Crystallography Open Database[37]). Pattern were obtained directly from the papers, which were rotated during the measurement. The crystallite size regime was estimated from the diffractograms via the Scherrer equation after performing a Rachinger correction, assuming a shape factor of 0.94 and calculating the full width at half maximum from Pseudo-Voigt reflection fits according to Olivero and Longbothum,[38] accounting for the instrumental line broadening.

*Random laser setup:* A Q-switched Nd:Yag laser operating at 532nm, 4 ns pulse duration and 10 Hz repetition rate is used as pump; a motorized rotating λ/2 plate with a polarized beam splitter are used for automatic energy variation detected by a power meter. The beam is focalized with a cylindrical lens to a diameter waist $W_P$ of the gaussian profile along Y axis variable in the range of 16-36 μm inside the sample. The emission is magnified with 100X magnification by an objective with numerical aperture 0.4 and splitted into two parts for both fluorescence imaging on a CCD camera and spectral detection. A filter is used for removing residual pump light. A fiber is mounted on motorized stages to scan the whole emission with a resolution of 0.5 μm and connected to the spectrograph equipped with electrically cooled CCD array detector; we use gratings density of 1800 mm$^{-1}$, corresponding to 0.07 nm spectral resolution. The same emission is imaged on the camera and a calibration tool is used to have the exact correspondence between each position of the fiber and each point of the CCD display. All spectra are the average of 100 single shots.

**Acknowledgements**
We thank MD Deen Islam for technical support and Marco Leonetti for fruitful discussions. We acknowledge support from the ERC project VANGUARD (grant number 664782) and the Templeton Foundation (grant number 58277).

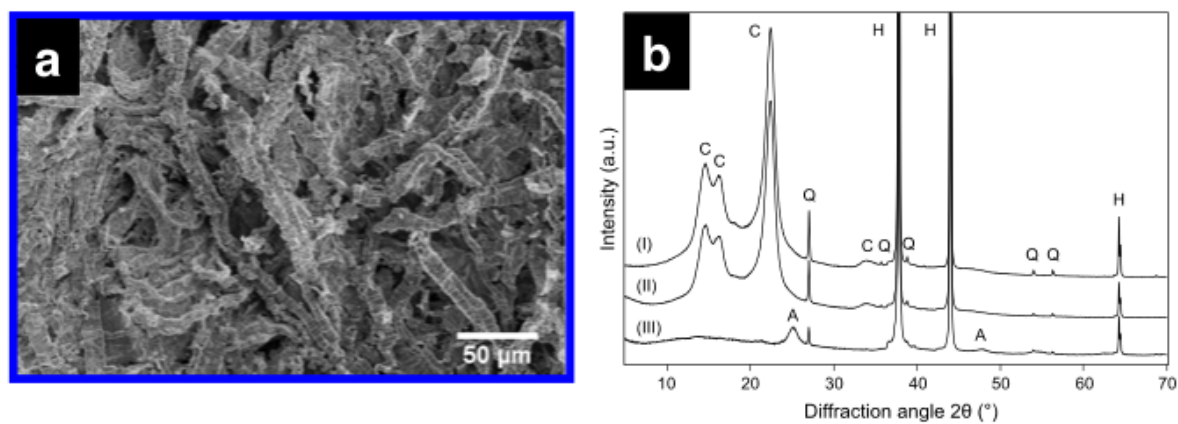

**Figure 1.** Biotemplated ceramic paper characterization. a) Scanning electron micrographs revealing the uniform replication of the cellulose fibers on the micrometer scale. b) X-ray diffractograms of (I) the native paper template, which was then (II) infiltrated and (III) calcined at 500 °C. Reactions can be assigned to (C) cellulose, (Q) quartz, (A) anatase, and (H) holder, respectively.

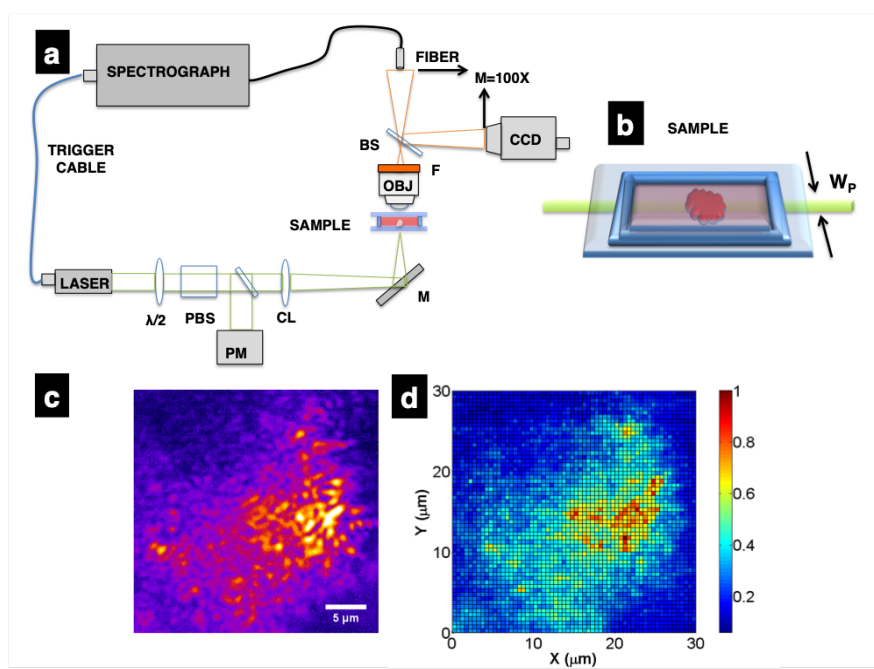

**Figure 2.** Imaging and intensity scanning of the emission. a) Sketch of the experimental setup: a λ/2 plate with a polarized beam splitter PBS are used for variation detected by a power meter PM; the beam is focalized by a cylindrical lens CL inside the sample and the emission is magnified by the objective OBJ and splitted by a beam splitter BS for both uorescence imaging on a charge coupled device (CCD ) and fiber scanning spectrography. A filter F is used for removing residual pump light. b) Scheme of the sample illuminated by the laser beam. c) Fluorescence image of a section of dyed TiO2 calcinated paper pumped at $E_P$=6.4 μJ energy. d) 30 μm X 30 μm spatial mapping of the emission from the system imaged in panel c. The values are obtained by integrating the spectral intensity point by point along X and Y.

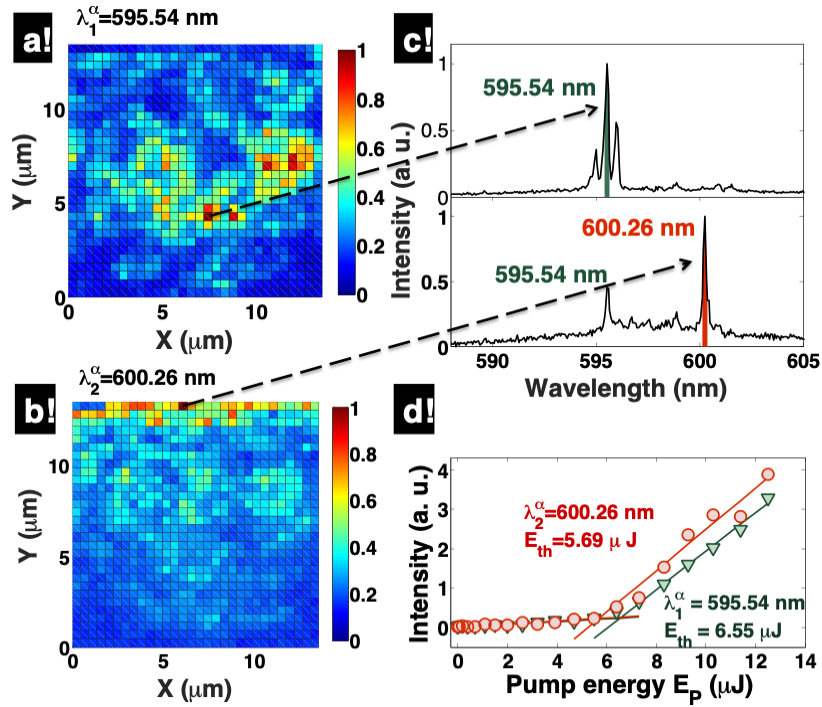

**Figure 3.** Mapping of distinct RL modes. a,b) 13.5 μm X 13.5μ m normalized intensity maps of two sharp resonances at $\lambda_1^\alpha$=594.54 nm (a) and $\lambda_2^\alpha$ 2=600.26 (b) for sample α. The values are obtained by integrating the intensity of single peaks, point by point along X and Y. c) Representative normalized spectra of the two modes taken at the indicated positions. d) Peak intensity of single resonances versus pump energy $E_P$ showing typical RL behavior; $E_{th}$ is the threshold of the laser modes. Solid lined through data are the linear fitting curves.

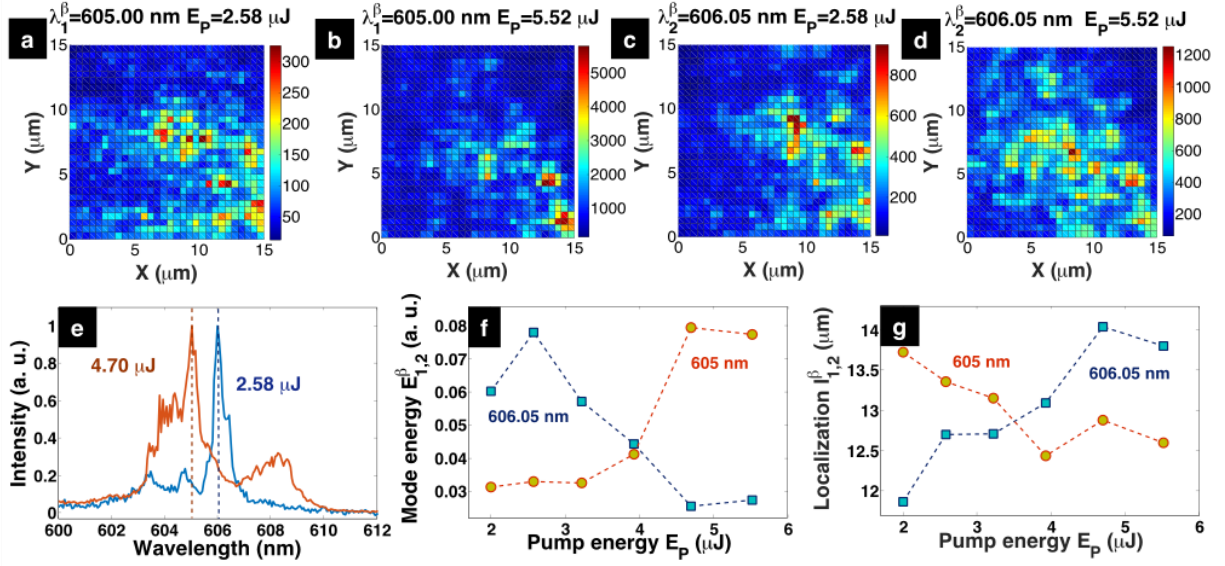

**Figure 4.** Observation of mode competition and resonance modulation. a-d) Intensity maps of two modes $\lambda_1^\beta$ and $\lambda_2^\beta$ at two different pump energy $E_P$ for sample β. The values are obtained by integrating the intensity of single peaks, point by point along X and Y. e) Representative normalized spectra at two different pump energy showing the transfer of energy between the resonances. f-g) Energy $E^\beta_{1,2}$ (f) and localization length $l^\beta_{1,2}$ (g) of the two modes vs. input energy $E_P$. Error bars are within the symbols.

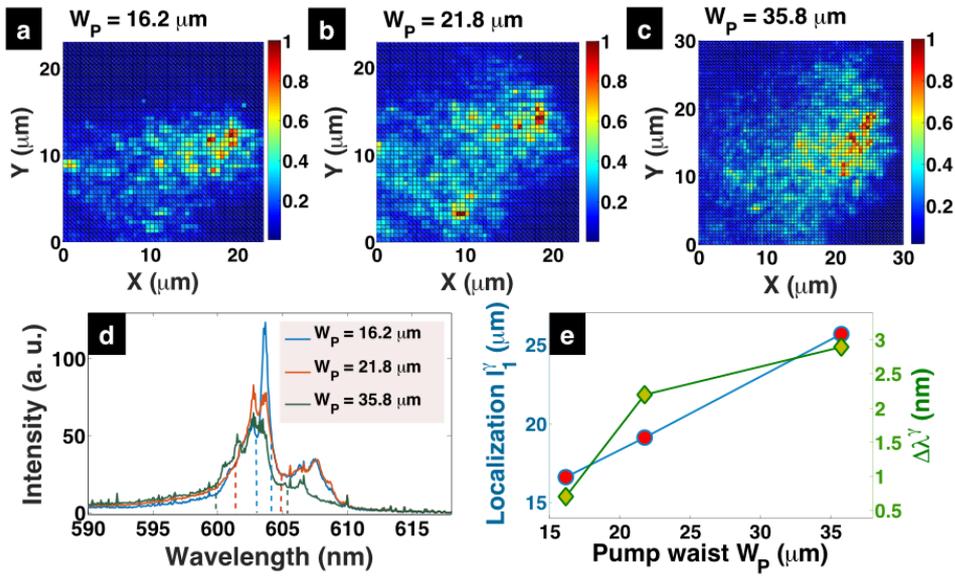

**Figure 5.** Control of spatial and temporal coherence. a-c) Normalized intensity maps of one mode in sample γ at three different pumping beam width $W_P$. The intensity in each pixel along X and Y is the area of one peak that broadens for increasing $W_P$. d) Spectra at different $W_P$, showing change of line shape and diverse spectral configuration. e) Localization length $l_1^\gamma$ and spectral line width $\Delta\lambda^\gamma$ vs. pump laser width $W_P$. Error bars on both axes are within the symbols.

**TABLE OF CONTENT**

Biologically inspired photonic structures are the key for technological advances and for miniaturized lasers. The proposed devices are few modes random lasers in dye-doped titania with the disordered structure of standard paper. Highly sharp electromagnetic resonances with unexpected broad spatial extension are reported. The modes interact and compete on wide length scales enabling to control precisely the device performance by acting on the pumping.

Keywords: Random lasers, biomimetic photonics, nonlinear optics in complex systems, coherence control

*N. Ghofraniha\*, L. La Volpe, D. Van Opdenbosch, C. Zollfrank, C. Conti*

***Biomimetic random lasers with tunable spatial and temporal coherence***

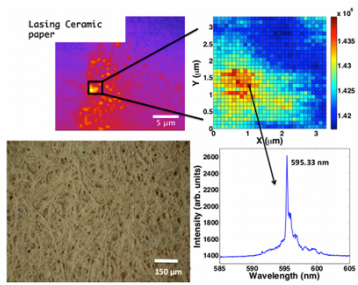